\documentstyle[12pt]{article}

\def\o{\over}
\def\b{\begin{equation}}
\def\e{\end{equation}}

\def\kpnn{$K^+\to\pi^+\nu\bar\nu$\ }
\def\kpn{K^+\to\pi^+\nu\bar\nu}

\begin{document}
\thispagestyle{empty}
\begin{flushright}
 MPI-PhT/94-3 \\
 January 1994
\end{flushright}
\vskip1truecm
\centerline{\Large\bf \kpnn Beyond Leading Logs and $V_{td}$
  \footnote[1]{\noindent
 Talk given at the International Europhysics Conference on
High Energy Physics, Marseille, July 22-28, 1993 }}
\vskip2truecm
\centerline{\sc Gerhard Buchalla}
\bigskip
\centerline{\sl Max-Planck-Institut f\"ur Physik}
\centerline{\sl  -- Werner-Heisenberg-Institut --}
\centerline{\sl F\"ohringer Ring 6, D-80805 M\"unchen, Germany}
\vskip2.5truecm
\centerline{\bf Abstract}
The calculation of QCD corrections beyond the leading logarithmic
approximation for the short-distance dominated rare decay \kpnn is
summarized. This analysis requires the complete ${\cal O}(\alpha_s)$
corrections to the top contribution to all orders in the top-quark mass
and a two-loop renormalization group (RG) calculation for the charm
contribution. The inclusion of these QCD corrections reduces considerably
the theoretical uncertainty in calculating $B(\kpn)$. Implications for
the determination of $V_{td}$ from this decay are discussed.
\vfill
\newpage

\pagenumbering{arabic}

\section{Introduction}

One of the most interesting possibilities for future tests of the
standard model (SM) by
investigating rare decay phenomena is provided by the
process $\kpn$. This decay mode has attracted the attention of many
authors in the past ([1--7] and refs. cited therein) due to several
features which make it a unique decay to study. \kpnn is a
flavor-changing neutral current (FCNC) transition, induced in the SM
at the one-loop level through the Z-penguin- and box-type dia\-grams
shown
in fig. 1, where also QCD radiative corrections have to be added.
Being a SM loop effect, the \kpnn transition probes
flavordynamics at the quantum level. This in turn allows for an indirect
test of physics at high energy scales in a low energy process. In
particular this decay is sensitive to the mass ($m_t$) and CKM couplings
($V_{ts}$, $V_{td}$) of the top quark, parameters related to the
symmetry breaking sector of the SM.\hfill\break
In addition, possible long-distance contributions to
\kpnn can be shown to be
negligibly small [8, 9]. Since the decay is semileptonic, the hadronic
matrix element involved is just the matrix element of a current and
can be extracted by isospin symmetry from the leading decay
$K^+\to\pi^0e^+\nu$. Together this implies that \kpnn is practically
short-distance dominated and QCD effects are entirely perturbative.
\hfill\break
Thus, besides the phenomenological interest of the underlying
quark-level QFD processes, \kpnn has the particular advantage of
being clean theoretically, that is, being afflicted with practically no
further theo\-retical uncertainties than those inevitably related to
QCD perturbation theory. Since, as we will see, the theoretical
uncertainty in the leading log approximation is quite substantial, the
extension of the QCD analysis to next-to-leading order is therefore
clearly desirable.\hfill\break
The relevant low energy effective hamiltonian induced by the diagrams
of fig. 1 can be written as
\b\label{heff}
{\cal H}_{eff}={G_F\o\sqrt 2}{\alpha\o 2\pi\sin^2\Theta_W}
 \left( V^*_{ts}V_{td} X(x_t)+V^*_{cs}V_{cd} X_{ch}\right)
 (\bar sd)_{V-A}(\bar\nu\nu)_{V-A}
\e
Here $X(x_t)$ (where $x_t=m^2_t/M^2_W$) and $X_{ch}$ are functions
re\-pre\-sen\-ting the con\-tri\-bu\-tions with virtual
top and charm. In writing
(\ref{heff}) the up quark contribution has been eliminated by means
of the unitarity of the CKM matrix. Although the top quark gives
a somewhat bigger contribution than charm, both contributions are of
comparable size since the effect of the large top mass $m_t\gg m_c$ is
compensated through the CKM suppression $|V^*_{ts}V_{td}|\ll
|V^*_{cs}V_{cd}|$.\hfill\break
The current experimental upper limit for the branching ratio is
$B(\kpn)\leq 5.2\cdot 10^{-9}$ [10]. There are ongoing experimental
efforts to close the gap to the SM expectation of about
(0.5 -- 6) $\cdot 10^{-10}$ in the coming years [11].

\section{QCD Corrections}

The formal basis of the present analysis is the operator product
expansion. It allows one to calculate the Wilson coefficients
$X(x_t)$, $X_{ch}$ which contain the information on the
short-distance physics and give the coupling strength of the local
four fermion operator $(\bar sd)_{V-A}(\bar\nu\nu)_{V-A}$, the
interaction term in the effective theory (\ref{heff}). Since this
operator has no anomalous dimension, the functions $X(x_t)$, $X_{ch}$
are independent of the renormalization scale $\mu$. However this is true
only up to terms of the neglected order in QCD perturbation theory.
The resulting residual $\mu$-dependence constitutes the intrinsic
theoretical uncertainty of any perturbative QCD calculation. This is
well known, but has not been discussed previously for the case at hand.
It turns out that in the leading logarithmic approximation (LLA) used
so far this uncertainty is quite big but it can be reduced
significantly at next-to-leading order (NLLA). The QCD analysis is
described in detail in [7]. Here we will content ourselves with briefly
summarizing the most important aspects.\hfill\break
The structure of the calculation is different for the top- and for the
charm-sector due to the fact that $m_t={\cal O}(M_W)$, but $m_c\ll M_W$.
This implies that the top-function $X(x_t)$ has to be calculated to
all orders in $m_t/M_W$, but usual perturbation theory can be applied
to analyze QCD effects. By contrast a RG calculation has to be performed
for the charm function $X_{ch}$ because large logs $\ln m_c/M_W$
appear, but it is sufficient to work to order $x_c=m^2_c/M^2_W$ in the
mass ratio. To leading order no QCD corrections are needed for the
top-sector (no large logs) and the leading log terms
$x_c \alpha^n_s \ln^{n+1} x_c$ are summed to all orders in $X_{ch}$.
The next-to-leading order calculation performed in [7] then includes
${\cal O}(\alpha_s)$ corrections in the former case and resums the
next-to-leading logs $x_c \alpha^n_s \ln^n x_c$ in the latter. In going
from leading to next-to-leading order the dependence on the
renormalization scale $\mu_t={\cal O}(m_t)$ ($\mu_c={\cal O}(m_c)$ for
charm) is reduced from ${\cal O}(\alpha_s)$ to ${\cal O}(\alpha^2_s)$
for $X(x_t)$ and from ${\cal O}(x_c)$ to ${\cal O}(x_c \alpha_s)$ for
$X_{ch}$. Numerically this amounts to a reduction from 10\% to 1\%
for $X(x_t)$ ($m_t(m_t)=150GeV$) and from 50\% to less than 20\% for
$X_{ch}$ when the scales are varied in the ranges
$100GeV\leq\mu_t\leq 300GeV$ and $1GeV\leq\mu_c\leq 3GeV$
respectively. Since the relevant scale is lower in the charm sector,
the QCD coupling is stronger and bigger uncertainties remain.

\section{Phenomenological Implications}

For typical values of the necessary input parameters
($m_t(m_t)=150GeV$,\hfill\break $m_c(m_c)=1.3GeV$,
$\Lambda_{\overline{MS}}=0.25GeV$,
Wolfenstein-parameters: $A=0.89$, $\varrho=0.08$, $\eta=0.44$)
one finds, with $\mu_t=m_t$ and $\mu_c=m_c$, $B(\kpn)=10^{-10}$
(three neutrino flavors). The theoretical uncertainty in this prediction
due to the $\mu$-dependences discussed above is reduced from 41\% to
11\% when next-to-leading corrections are included. The game can be
reversed and $|V_{td}|$ can be extracted from a measured branching
ratio. Assuming $B(\kpn)=10^{-10}$ we find $|V_{td}|=0.01$ with an
uncertainty of 30\% in LLA but only 7\% in NLLA.\hfill\break
We have concentrated on the issue of the scale ambiguity, which
represents the intrinsic theoretical uncertainty and limits the
accuracy that may be ultimately achieved in any phenomenological
application of $\kpn$. At present the most important uncertainties come
still from the fact that $B(\kpn)$ is not yet measured and $m_t$ is
only poorly known. However this will hopefully improve in the future.
Then, ultimately, the gain in accuracy by a factor of four in the
theoretical prediction in NLLA
\begin{itemize}
\item will enable an improved determination of $|V_{td}|$,
\item may help in particular to resolve the two-fold ambiguity
   in the CKM-phase $\delta$ and will finally be
\item mandatory for a decisive test of the consistency of the
  SM picture of quark mixing.
\end{itemize}

\vskip 6.0mm

I am grateful to Andrzej Buras for his constant support and
encouragement and for a very pleasant collaboration.

\section*{References}

\noindent  [1]
A.J. Buras and M.K. Harlander, A Top Quark Story, {\it in} Heavy Flavors,
eds. A.J. Buras and M. Lindner, World Scientific (1992), p.58
\newline
\noindent  [2]
L. Littenberg and G. Valencia, BNL-48631, {\it to appear in}
{\sl Ann. Rev. Nucl. Part. Science}, Volume 43
\newline
\noindent [3]
T. Inami and C.S  Lim, {\sl Prog. Theor. Phys.} {\bf 65}
(1981) 297;1772
\newline
\noindent [4]
J. Ellis and J.S. Hagelin, {\sl Nucl. Phys.} {\bf B 217} (1983) 189;
V.A. Novikov, M.A. Shifman, A.I. Vainshtein and V.I. Zakharov,
{\sl Phys. Rev.} {\bf D 16} (1977) 223
\newline
\noindent [5]
C.O. Dib, I. Dunietz and F.J. Gilman, {\sl Mod. Phys. Lett.} {\bf A 6}
(1991) 3573
\newline
\noindent [6]
G. Buchalla, A.J. Buras and M.K. Harlander, {\sl Nucl. Phys.} {\bf B 349}
(1991) 1; G. Buchalla, Diploma-Thesis (1990), unpublished
\newline
\noindent [7]
G. Buchalla and A.J. Buras, {\sl Nucl. Phys.} {\bf B 398} (1993) 285,
{\bf B 400} (1993) 225; MPI-Ph/93-44
{\it to appear in} {\sl Nucl. Phys. B}
\newline
\noindent [8]
D. Rein and L.M. Sehgal, {\sl Phys. Rev.} {\bf D 39} (1989) 3325
\newline
\noindent [9]
J.S. Hagelin and L.S. Littenberg, {\sl Prog. Part. Nucl. Phys.}
{\bf 23} (1989) 1
\newline
\noindent [10]
M.S. Atiya et al., {\sl Phys. Rev. Lett.} {\bf 64} (1990) 21;
BNL-48066, BNL-48091, {\it subm. to} {\sl Phys. Rev. Lett.}
\newline
\noindent [11]
Y. Kuno, KEK Preprint 92-128
\vfill
\end{document}